

                                                        %
\font\smb=cmss10                                        %
\font\secrmb=cmr10 scaled 1440                          %
\font\rmb=cmr10 scaled 2074                             %
                        %
                                                        %
\font\euler=eufm10                                      %
\newfam\eufmfam                                         %
\textfont\eufmfam=\euler                                %
\def\g{{\ifmmode\frak{g}\else\euler g\fi}}              %
\def\h{{\ifmmode\frak{h}\else\euler h\fi}}              %
                                                        %

\input epsf.tex

\def\plq{{\mathop{{\vcenter{\hrule height .5pt
    \hbox{\vrule width .5pt height 4pt \kern 4pt
    \vrule width .5pt} \hrule height .5pt}}}}}
\def\la{\raise.16ex\hbox{$\langle$}}
\def\ra{\raise.16ex\hbox{$\rangle$}}
\def\d{\partial}

\def\Z{ \hbox{\smb Z} \kern-.4em \hbox{\smb Z} }
\def\R{ \hbox{I} \kern-0.47em \hbox{ \smb R} }
\def\E{{\bf E}~}
\def\tr{{\rm Tr}~}
\def\frak#1{{\fam\eufmfam\relax#1}}
\def\half{{1\over 2}}

\def\leaderfill{\leaders\hbox to 1em{\hss.\hss}\hfill}
\def\qstretch{\qquad\qquad\qquad\qquad\qquad\qquad\qquad\qquad\qquad\qquad}

\global\newcount\secno \global\secno=0
\global\newcount\meqno \global\meqno=1
\global\newcount\figno \global\figno=1
\newwrite\eqmac
\def\eqn#1{
        \ifnum\secno>0
            \eqno(\the\secno.\the\meqno)\xdef#1{\the\secno.\the\meqno}
          \else\ifnum\appno>0
            \eqno(\romappno.\the\meqno)\xdef#1{\romappno.\the\meqno}
          \else
            \eqno(\the\meqno)\xdef#1{\the\meqno}
          \fi
        \fi
\global\advance\meqno by1 }
\newwrite\figmac
\def\fig#1{\ifnum\secno>0
             \the\figno\xdef#1{\the\figno}
           \fi
        \global\advance\figno by1 }

\global\newcount\refno
\global\refno=1 \newwrite\reffile
\newwrite\refmac
\newlinechar=`\^^J
\def\ref#1#2{\the\refno\nref#1{#2}}
\def\nref#1#2{\xdef#1{\the\refno}
\ifnum\refno=1\immediate\openout\reffile=refs.tmp\fi
\immediate\write\reffile{
     \noexpand\item{[\noexpand#1]\ }#2\noexpand\nobreak.}
     \immediate\write\refmac{\def\noexpand#1{\the\refno}}
   \global\advance\refno by1}
\def\semi{;\hfil\noexpand\break ^^J}
\def\refn#1#2{\nref#1{#2}}

\def\cmp#1#2#3{{\it Comm. Math. Phys.} {\bf {#1}} (19{#2}) #3}
\def\ijmp#1#2#3{{\it Int. J. Mod. Phys.} {\bf {#1}} (19{#2}) #3}

\def\pl#1#2#3{{\it Phys. Lett.} {\bf {#1}B} (19{#2}) #3}
\def\np#1#2#3{{\it Nucl. Phys.} {\bf B{#1}} (19{#2}) #3}
\def\pr#1#2#3{{\it Phys. Rev.} {\bf {#1}} (19{#2}) #3} 

\def\prD#1#2#3{{\it Phys. Rev.} {\bf D{#1}} (19{#2}) #3}
\def\prl#1#2#3{{\it Phys. Rev. Lett.} {\bf #1} (19{#2}) #3}
\def\rmp#1#2#3{{\it Rev. Mod. Phys.} {\bf {#1}} (19{#2}) #3}
\def\ann#1#2#3{{\it Ann. Phys.} {\bf {#1}} (19{#2}) #3}
\def\prp#1#2#3{{\it Phys. Rep.} {\bf {#1}C} (19{#2}) #3}

\refn\Wilson{K. Wilson, \prD{14}{74}{2455}}
\refn\Wittena{E. Witten, \cmp{141}{91}{153}}
\refn\Wittenb{E. Witten, J. Geom. Phys. {\bf 9} (1992) 303,
      {\tt  hep-th/9204083}}
\refn\Danny{D. Birmingham, M. Blau, M. Rakowski, and G. Thompson,
     \prp{209}{91}{129}}
\refn\Rusakov{B. Rusakov, Mod. Phys. Lett. {\bf A 5} (1990) 693}
\refn\Wheater{J.F. Wheater, \pl{264}{91}{161}}
\refn\BlauThompson{M. Blau and G. Thompson, \ijmp{A 7}{91}{3781}}
\refn\DH{J.J. Duistermaat and G.J. Heckman, Invent. Math. {\bf 69}
      (1982) 259}
\refn\Rajeev{S.G. Rajeev, \pl{212}{88}{203}}
\refn\HH{J.E. Hetrick and Y. Hosotani, \pl{230}{89}{88}}
\refn\GrossTaylor{D. Gross and W. Taylor, LBL preprint {\it
      LBL-33458}, {\tt hep-th/9301068}}
\refn\Minahan{J. Minanhan, \prD{47}{93}{3430},
      {\tt hep-th/9301003}}
\refn\tHooftN{G. 't Hooft, \np{74}{72}{461}}
\refn\KK{V. Kazakov and I. Kostov, \np{176}{80}{199}}
\refn\JHAmstlat{J.E. Hetrick, Nucl. Phys. {\bf B} (Proc. Suppl.) {\bf
      30} (1993) 228}
\refn\LSb{E. Langmann and G. Semenoff, \pl{303}{93}{303},
      {\tt hep-th/9212038}}
\refn\Sundemeyer{K. Sundemeyer, {\it Constrained Dynamics},
      Springer, 1982}
\refn\PS{A. Pressely and G. Segal, {\it Loop Groups}, Oxford Clarendon
      Press, 1986}
\refn\Mickelsson{J. Mickelsson, \pl{242}{90}{217}}
\refn\Yutaka{Y. Hosotani, \ann{190}{89}{233}}
\refn\GmodG{M. Blau and G. Thompson, ITCP preprint {\it IC/93/83},
      {\it Derivation of the Verlinde Formula from Chern-Simons Theory
      and the $G/G$ Model}, {\tt hep-th/9305010}}
\refn\DowkerCone{J.S. Dowker, J. Phys {\bf A10} (1977) 115}
\refn\tHooftFlux{G. 't Hooft, \np{153}{79}{141}}
\refn\Pierre{P. van Baal, \cmp{85}{82}{529}}
\refn\Mark{M. Burgess, A. McLachlan, and D.J. Toms, U. Oslo
      preprint 92-01 {\it Dynamics of Magnetic Fields in Maxwell,
      Yang-Mills and Chern-Simons Theories on the Torus}}
\refn\LSa{E. Langmann and G. Semenoff, \prl{296}{92}{117},
      {\tt hep-th/9210011}}
\refn\KS{J. Kogut and L. Susskind, \prD{11}{75}{395}}
\refn\Kogut{J. Kogut, \rmp{55}{83}{775}}
\refn\Creutz{M. Creutz, {\it Quarks, Gluons, and Lattices},
      Cambridge Univ. Press, 1988}
\refn\Jan{J. Smit,  lecture notes on lattice gauge theory, unpublished}
\refn\DowkerG{J.S. Dowker, \ann{62}{71}{361}}
\refn\Vilenkin{N. Ja. Vilenkin, {\it Fonctions Sp\'eciales et
      Th\'eorie de la Repr\'esentation des Groupes}, Dunod, 1969}
\refn\Berezin{F.A. Berezin, {\it Amer. Math. Soc. Transl.} {\bf 21}
      (2) (1962) 239}
\refn\Helgason{S. Helgason, {\it Groups and Geometric Analysis,
      Integral Geometry, Invariant Differential Operators and
      Spherical Functions}, Academic Press, 1984}
\refn\Schulman{L. Schulman, \pr{176}{68}{1558}}
\refn\Cahn{R. Cahn, {\it Semi-Simple Lie Algebra and Their
      Representations}, Benjamin/Cummings Pub. Co., 1984}
\refn\Ruse{H.S. Ruse, A.G. Walker, and T.J. Willmore,
      {\it Harmonic Spaces}, Edizioni Cremonese, Roma, 1961}
\refn\Dowkerpath{J.S. Dowker, {\it J. Phys.} {\bf A3} (1970) 451}
\refn\Stone{M. Stone, \np{314}{89}{557}}
\refn\Renata{R. Loll, \prD{41}{90}{3785}}
\refn\KalauBRST{W. Kalau, \np{}{93}{}, {\tt hep-th/9209035}}
\refn\JS{A. Jevicki and B. Sakita, \prD{22}{80}{467}}

\magnification=1200
\baselineskip=12pt plus 1pt

\nopagenumbers

\rightline{hep-th/9305020}
\rightline{UvA-ITFA 93-15}
\vskip 3mm
\rightline{May 1993}

\baselineskip=20pt plus 1pt
\vskip 1.5cm

\centerline{\rmb Canonical Quantization of}
\centerline{\rmb Two Dimensional Gauge Fields}
\vskip 0.75cm

\centerline{\bf James E. Hetrick}
\centerline{\tt hetrick@phys.uva.nl}
\smallskip
\baselineskip=12pt plus 1pt
\centerline{\it Institute for Theoretical Physics}
\centerline{\it University of Amsterdam}
\centerline{\it Valckenierstraat 65}
\centerline{\it 1018-XE Amsterdam}
\centerline{\it The Netherlands}

\vskip 1.5cm
\baselineskip=12pt plus 1pt
\parindent 20pt
\centerline{\bf Abstract} \midinsert \narrower \frenchspacing

$SU(N)$ gauge fields on a cylindrical spacetime are canonically
quantized via two routes revealing almost equivalent but different
quantizations. After removing all continuous gauge degrees of freedom,
the canonical coordinate $A_\mu$ (in the Cartan subalgebra $\h$) is
quantized. The compact route, as in lattice gauge theory, quantizes the
Wilson loop $W$, projecting out gauge invariant wavefunctions on the
group manifold $G$. After a Casimir energy related to the
curvature of $SU(N)$ is added to the compact spectrum, it is seen to be
a subset of the non-compact spectrum. States of the two quantizations
with corresponding energy are shifted relative each other, such that
the ground state on $G$, $\chi_0(W)$, is the first excited state
$\Psi_1(A_\mu)$ on $\h$.  The ground state $\Psi_0(A_\mu)$ does not
appear in the character spectrum as its lift is not globally
defined on $G$. Implications for lattice gauge theory and the sum
over maps representation of two dimensional QCD are discussed.
\endinsert

\vfill\eject
\footline={\hss\tenrm\folio\hss}
\pageno=2

\leftline{\it Contents:}
\smallskip
\line{I. Introduction \leaderfill 2 \qstretch}
\line{II. The Topology of Phase Space \leaderfill 5 \qstretch}
\line{III. Quantization: \hfill \qstretch}
\line{~~A. Compact Formulation \leaderfill 9 \qstretch}
\line{~~B. Non-compact Formulation \leaderfill 16 \qstretch}
\line{IV. Spectral Equivalence \leaderfill 19 \qstretch}
\line{V. State Mapping \leaderfill 24 \qstretch}
\line{VI. Conclusions \leaderfill 27 \qstretch}
\leftline{~~~~~~~~~~~~~~~~~~---------$\infty$---------}

\parindent 30pt
\baselineskip=20pt
\secno=1  \meqno=1
\bigskip
\leftline{\secrmb I. Introduction}
\bigskip

Gauge theories are by nature overdetermined systems in which many
different field configurations are in fact physically equivalent, thus
the primary issue to be addressed in their quantization is the
treatment of the field's excessive degrees of freedom and the
identification of the equivalence classes under gauge transformations.

Discretely formulating a gauge theory on the lattice as done by Wilson
[\Wilson], introduces a novel solution to the gauge fixing problem, that
is to say, it becomes a finite problem and can be addressed implicitly
at the expense of exploring many extra dimensions in phase space. In the
continuum the set of all allowed gauge transformations is an infinite
dimensional space, the volume of which is delicately factored out of
the partition function by the Fadeev-Popov Jacobian. On the lattice,
the volume of gauge space is finite, and gauge degrees of freedom can
be left in the path integral measure. They are divided out by the
normalization when only gauge invariant quantities are computed.

The Wilson formulation of lattice gauge theory of course alters
the form of the theory in a further very significant way; not only
does it make the configuration space finite dimensional, but also
compact. The (non-compact) gauge field
$A_\mu$ becomes the (compact) link $U_\mu$,
$$
A_\mu(x) \in \g~ \rightarrow~U_\mu(x) =
\exp\{i\int_x^{x+a}dy\cdot A_\mu(y)\} \in G
\eqn\AtoU
$$
where $G$ is the gauge group, and $\g$ its Lie algebra. In the
classical continuum limit as the lattice spacing $a$ goes to zero,
this distinction should disappear since the group manifold $G$ looks
more and more like its tangent space $\g$ as the link $U_\mu$ is
limited to a small region near the identity. As we shall discover
however, the topological distinction remains for quantities which
depend on the global structure of phase space such as the zero modes
of the Hamiltonian.

We demonstrate this in two spacetime dimensions since there we can
completely eliminate the gauge degrees of freedom and explicitly solve
the model in full, a luxury not available in higher dimensions. The
results may be applicable to compact theories in higher dimensions,
since they depend only on the relation between differential operators
under the compactification of the Lie algebra $\g$ to $G$ via the
exponential map, however the Hamiltonian in a higher dimensional
theory is much more complicatied, as is the gauge orbit structure.
Furthermore the physical degrees of freedom are not all {\it radial}
coordinates of $\g$ or $G$ in higher dimensions, an aspect which is
crucial here.

We will use the canonical formalism, since our focus is the
Hamiltonian as a differential operator. It is well known that planar
Yang-Mill is trivial, since the gauge fields can be transformed to
zero throughout spacetime.  Another reason planar Yang-Mills is
trivial is that two dimensional Yang-Mills (2DYM) is a form of
topological field theory
[\Wittena,\Danny,\Rusakov,\Wheater,\BlauThompson], whose excitations
depend only on the topology of the underlying spacetime.  Thus we use
a cylindrical spacetime, the only topologically non-trivial two
dimensional manifold, with canonical time.

Investigations of gauge fields on arbitrary Riemann surfaces have
begun in earnest and various facets were recently collected in two
very probing works by Witten [\Wittena,\Wittenb]. The first of these
papers examines 2DYM in several ways: from the explicit lattice
formulation, as the limiting cases of Chern-Simons and conformal field
theories, and its relation to the theory of Reidmeister torsion.  In
the second paper, Witten re-examines 2DYM by generalizing the
Duistermaat-Heckman integration formula [\DH] to a non-Abelian form,
yielding the Yang-Mills partition function as a polynomial in the
coupling constant, ie. as a sum over critical points of phase space
using the action as a Morse functional.

The focus of the present paper is a minute difference between these
papers already addressed in [\Wittenb], namely that the spectrum in
[\Wittena], obtained in the lattice formulation of 2DYM is half the
quadratic Casimir $C_2/2$ of $SU(N)$, whereas in [\Wittenb] it is
$C_2/2 + t$, where $t$ is a ``lower order Casimir'' to be determined
by the regularization of the theory and its connection to the
equivalent topological $BF$ theory.

This difference in spectra was previously seen in two canonical
continuum solutions of $SU(N)$ on a cylinder, by Rajeev [\Rajeev] where
the quantization is done on the group manifold, and by Hosotani and
the author [\HH], where the quantization is done in the algebra.  We
will see that the difference corresponds exactly to Witten's $t$
parameter, and is related to the mapping of the Hamiltonian as a
differential operator from the algebra to the group, via the
exponential map. More interesting than this constant rescaling of
vacuum energy, is a corresponding shift of states between the two
quantizations such that the ground states are different.  Thus we
arrive at two inequivalent quantizations of the theory, whose
excitations are in one-to-one correspondence, except for an extra set
of lower energy states which appear in the non-compact quantization.

Recently Gross and Taylor [\GrossTaylor] and Minahan [\Minahan], have
shown that the partition function of 2DYM (as $N \rightarrow \infty$)
is given by the sum over maps from compact genus $g$ worldsheets to
the two dimensional target spacetime, renewing interest in the $1/N$
and string representations of QCD [\tHooftN, \KK]. An
interesting feature of this expansion of the partition function,
relating representation theory of Lie groups to the classification of
maps of surfaces, is that degenerate maps in which the worldsheet is
mapped to a single point of the target space are, somewhat
mysteriously, absent.

In the canonical approach the nature of the excitations as Fourier
modes on the maximal torus of $G$ is made clear and we find that the
two different quantizations are essentially even and odd choices for
these modes. We will see that compact quantization naturally picks out
the odd modes, precluding a constant wavefunction on the group
algebra. We make conjectures about the relationship between this state
and the zero winding maps, and about the general analogies between the
canonical Fourier states and smooth maps in the conclusions.

The paper is organized as follows: in the next section we examine the
general features of the gauge orbit and phase space of the theory,
identifying explicitly the configuration space and its topology. We
find that the residual Gribov copies play a crucial role in creating
this topology [\JHAmstlat,\LSb]. We then study the theory canonically
in the diagonal-Coulomb gauge, computing the Hamiltonian, and
quantizing the theory in the group algebra.

We review the lattice method of quantization on the group manifold,
and compare the results.  This comparison leads us to the relationship
between Laplacians on group manifolds and their algebr\ae. With this
relationship in hand we examine the mapping of states in the two
quantizations and see that the curvature of $G$ induces a shift
between corresponding states. Finally we discuss the implications of
inequivalent quantizations of this theory.

\secno=2  \meqno=1
\vskip 1cm
\leftline{\secrmb II. The Structure of Phase Space}
\bigskip

Before gauge fixing and quantizing the specific model,
we review the classical phase space structure of Yang-Mills fields
momentarily to put various features in context [\Sundemeyer].

{}From the Yang-Mills Lagrangian
${\cal L} = {1\over 2} {\rm Tr} F_{\mu\nu}F^{\mu\nu}$  we have the
canonical momenta to $A_\mu^a$
$$
{\delta {\cal L}\over \delta \d_0 A^\mu} \equiv \Pi_\mu^a = F_{0\mu}^a .
\eqn\canmom
$$
In the unreduced phase space $\Gamma$, $\Pi_0^a$ vanishes
identically providing $N^2-1$ functional constraints
$$
\phi_p\big[A_\mu^a(x),\Pi_\mu^a(x)\big] = \Pi_0^a = 0
\eqn\phip
$$
and defining the primary constraint surface $\Gamma_c$.
Further reductions of this surface come from the secondary
constraints related to the gauge symmetries.

The Hamiltonian (setting
$\Pi_0 = 0$ explicitly) is
$$
H = {\rm Tr}\int dx~\big[ \Pi^1 \dot A_1 - \half \Pi^1\Pi_1 \big].
\eqn\dotham
$$
{}From $\Pi_1 \equiv \E = \dot A_1 - \d_1 A_0 + ig[A_0, A_1]$,
we have the Hamiltonian in terms of the
coordinates $(A_\mu^a(x),\Pi_\mu^a(x))$ on $\Gamma_c$
$$
H = {\rm Tr}\int dx~\big[\half {\E}^2 - A_0 D\cdot\E \big]
\eqn\ham
$$
which shows $A_0^a$ to be a non-dynamical Lagrange multiplier for
Gauss' law
$$ D \cdot {\E} = \d_1 E^1 + ig [A_1, E^1] = \phi_s^{(a)} \approx 0,
\eqn\gausslaw
$$
which are secondary constraints stemming from the fact that
the primary constraints above must be time independent,
$\{\Pi_0, H\} = D \cdot {\bf E} \approx 0$.
We now write the extended Hamiltonian, including the constraints
and their multipliers,
$$\eqalign{
H_{ext} =& ~H + {\rm Tr}\int dx~\big[\omega_p(x)\phi_p +
\omega_s(x)\phi_s(x)\big]\cr
=& ~{\rm Tr}\int dx~\big[\half {\bf E}^2 + \omega_p \Pi_0 +
(\omega_s - A_0) D \cdot {\E} \big].\cr}
\eqn\Hext
$$

Further we see that $\phi_p$ and $\phi_s$ generate infinitesimal
gauge transformations of the coordinates $A_\mu$ on the surface of
constraint $\Gamma_c$, since
$$\eqalign{
\delta_\Omega A_\mu^a(x) =&  \int d^2y \Big[
\omega^b_p(y) \{A_\mu^a(x), \Pi_0^b(y)\} +
\omega^b_s(y) \{A_\mu^a(x), D_1\Pi^1_b(y)\} \Big]\cr
=& ~D_\mu \Omega^a(x)\cr
{\rm where~~}& \omega^a_p(x) = D_0 \Omega^a(x);~~ D_1 \omega_s^a(x)
= -D_1 \Omega^a(x).\cr}
\eqn\gausstrans
$$
Thus starting from a single configuration $(A_\mu,\Pi_\mu)$ on
$\Gamma_c$, and evolving in time under two Hamiltonians
$H_{ext}(\omega^\prime)$ and $H_{ext}(\omega^{\prime\prime})$, we
arrive at two different configurations $(A_\mu^\prime,\Pi_\mu^\prime)$
and $(A_\mu^{\prime\prime},\Pi_\mu^{\prime\prime})$, which are
equivalent up to a gauge transformation.

We thus have a
fibration of $\Gamma_c$ by the gauge orbits, ie. points which are
related to each other by gauge transformations
$$
A_\mu^\prime = \Omega A_\mu \Omega^{\dagger}
                    - {i\over g} \Omega \d_\mu \Omega^{\dagger}
\eqn\gaugetrans
$$
which are identified. The equivalence classes of these points (the
different orbits), classify the {\it reduced} phase space $\Gamma_r$
which are the true independent degrees of freedom which we wish to
quantize.

Usually phase space is the cotangent bundle $T^*Q$, of the
configuration manifold $Q$, so that identifying $Q$, $Q_c$, and the
reduced space $Q_r$ is the essential matter and with luck, $T^*Q_r$
just comes along for the ride. In our case $Q_r$ turns out to be an
orbifold so that $T^*Q_r$ is not defined at every point, however we
can quantize on the smooth manifold covering $Q_r$ and
implement the orbifold identifications in the Hilbert space of the
quantum theory.

The structure of $Q_c$ in a gauge theory is that of a fiber bundle
$$
\matrix{{\cal G}&\longrightarrow&Q_c\cr
           ~    &   ~       &~~\downarrow \pi\cr
           ~    &   ~       &Q_r\cr}
\eqn\bundles
$$
where $Q_c$ is the space of connections of an $SU(N)$ bundle over
the spatial manifold $X$, ${\cal G}$ is the group of allowed gauge
transformations on $X$, and $Q_r$ is the set of equivalence classes
of points in $Q_c$ under the action of ${\cal G}$.
In the two dimensional model, $Q_c$ is labeled by the fields
$A_1(x)$ which are maps of the circle $S^1$ into the Lie algebra $\g$
of $G$
$$
Q_c = L\g = \{A_1: S^1 \rightarrow \g\}
\eqn\Lgdef
$$ As such they are coordinates of a well known manifold $L\g$, the
Lie algebra of the loop group $LG$ [\PS].
Furthermore, $\cal G$, the space of
gauge transformations of $A_1(x)$, is isomorphic to the loop group $LG$
itself,
$$
{\cal G} = LG = \{\Omega(x): S^1 \rightarrow G\}
\eqn\LGdef
$$

To classify the orbits of $LG$ in the space $L\g$, ie. the physical
configuration space $Q_r$, consider the following representation of
$L\g$ in which each element is almost ``pure gauge''.  Let $0 < x \le
2\pi$ be the coordinate on $S^1$.  To each element $A \in L\g$, there
is an associated $G$-valued function $f:\R \rightarrow G$
which is the integral curve of the vector field $A(x)$ on $G$,
satisfying
$$
i\d_x f f^\dagger = A.
\eqn\deff
$$
Take the boundary condition $f(0) = 1$. Since $A(x+2\pi) = A(x)$, $f$
is periodic up to a constant element of $G$
$$
f(x+2\pi) = f(x) W_A.
\eqn\fW
$$
Under the action of $\Omega(x) \in LG$, $A$
transforms to
$$
\widetilde A = \Omega\cdot A
= \Omega A \Omega^\dagger + i \d_x \Omega \Omega^\dagger
\eqn\LGonLg
$$
hence $f$ goes into $\widetilde f$
$$
\widetilde f(x) = \Omega(x) f(x) \Omega^\dagger (0).
\eqn\ftilde
$$
Notice the trailing factor of $\Omega^\dagger(0)$ necessary to maintain
$\widetilde f(0) = 1$. The quasi-periodicity (\fW) of $f$, with
$\Omega(2\pi) = \Omega(0)$ implies that
$$
W_A \rightarrow \widetilde W_{\widetilde A}
= \Omega(0) W_A \Omega^\dagger(0).
\eqn\Wtilde
$$
We thus have a homomorphism between $L\g$ and the space of maps
$\{f: f(x+2\pi) = f(x)W\}$ for some $W \in G$, and can label
every $A \in L\g$ by $(f_A(x), W_A)$ .

Now for arbitrary $f(x)$ and
$\widetilde f(x)$ we can find a gauge transformation
$$
\gamma = \widetilde f \gamma(0) f^\dagger
\eqn\ArbLG
$$
in $LG$ (note that $\gamma(0)$ remains undetermined), such that
$$
\gamma \cdot (f(x), W) \rightarrow (\widetilde f(x), \widetilde W )
\eqn\lamshift
$$
only when $\widetilde W$ and $W$ lie in the same conjugacy class of
$G$; then there will there be a $\gamma(0)$ which can take
$W$ into $\widetilde W$.
Thus, the manifold of physically distinct configurations, the
orbits of $LG$ in $L\g$, is isomorphic to the space of conjugacy
classes of $G$ [\Rajeev,\HH,\LSb,\Mickelsson].

This space is an orbifold made by identifying points in the maximal
Abelian subgroup of $G$ (the maximal torus $T_G$), under the action of
the Weyl group $W_G$, a discrete set of transformations which permute
the diagonal elements. These correspond to reflections in the
hyperplanes through the origin and perpendicular to the roots of $G$,
hence the Weyl group is isomorphic to the permutation group of $N$
elements, $S_{N}$. Pure Yang-Mills theory is furthermore invariant
under constant gauge transformations which lie in the center of the
gauge group, hence the gauge symmetry is actually $SU(N)/\Z_N$. The
effect of the $\Z_N$ symmetry on $T_G$ is rather mild and simply
changes its periodicity to $2\pi/N$ as opposed to $2\pi$ in an $SU(N)$
theory. Therefore we have an exact identification of the topology of
the gauge orbit space of $SU(N)$ Yang-Mills theory on a cylinder as
the orbifold $T_G/S_N \sim T^{N-1}/{S_N}$.

\secno=3  \meqno=1
\vskip 1cm
\leftline{\secrmb III. Quantization}
\bigskip
\leftline{\bf A. Non-compact formulation}
\bigskip

In this section we quantize 2DYM on the group algebra $\g$ where
the theory is originally defined in terms of $A_\mu$.
For concreteness we specify the model as given by the Lagrangian and
gauge fields defined on a cylindrical spacetime of
circumference $L$ as follows:
$$
\eqalign{
{\cal L} =& -{1\over 2} {\rm Tr} F_{\mu\nu} F^{\mu\nu}\cr
F_{\mu\nu} = & \d_\mu A_\nu - \d_\nu A_\mu +ig[A_\mu, A_\nu]\cr
A_\mu(x,t) = &  A_\mu(x+L,t) = A_\mu^a \lambda_a;
\qquad {\rm su(N) = Span\{}\lambda_a\}\cr}
\eqn\model
$$
Note that the periodic boundary condition for the gauge field may be
taken without loss of generality on a cylinder since the $SU(N)$
bundle over $S^1$ is trivial [\Yutaka].
Under gauge transformations (\gaugetrans), periodicity remains
intact provided the gauge transformation satisfies the condition
$$
\Omega(x+L,t) = \Z_n \Omega(x,t)
\eqn\gaugeBC
$$
where $\Z_n$ is any element in the center of $SU(N)$.

As mentioned above we use the {\it diagonal-Coulomb} gauge [\HH]
defined as
$$\eqalign{
\d_1 A_1 &=  0 \cr
A_1(t)_{ij} &=  {1\over gL}\theta_i(t) \delta_{ij} \cr}
\eqn\diagCoulombdef
$$
which provides coordinates directly on the torus covering the reduced
configuration space $T^{N-1}/S_N$. It is easy to see that the gauge
transformation
$$
\Omega(x,t) = \Lambda(t) {\cal W}(x,t) V^\dagger (x,t)
\eqn\DiagCoulOmega
$$
where
$$\eqalign{
V(x,t) =&  {\cal P}\exp\big( -ig\int_0^x dy A_1(y,t)\big)\cr
{\cal W}(x,t) =&  \exp[-ixB(t)/L],\qquad \exp[-iB(t)] = V(L,t)\cr
\Lambda(t) \in G:~~\Lambda B \Lambda^\dagger
=&  \left(\matrix{\theta_1(t)&~&~\cr
{}~&\ddots&~\cr
{}~&~&\theta_N(t)}\right) \equiv \Theta(t),
\qquad \sum_{i=1}^N \theta_i(t) = 0\cr}
\eqn\DCOmegaExplicit
$$
takes $A_1(x,t)$ into $(1/gL) \Theta(t)$. We use
here a Weyl parameterization of the Cartan subalgebra $\h$
as it illuminates the toroidal
structure in a simple fashion. We will turn to other bases
later. In terms of the previous loop group decomposition in eq.
(\ArbLG), $V(x,t) = f_{A_1}$, ${\cal W}(x,t) = f_B$, and $\Lambda =
\gamma(0)$:
$$
\Omega = \Lambda f_B f^\dagger_{A_1} = \Lambda f_B \Lambda^\dagger
\Lambda f^\dagger_{A_1} = f_\Theta \gamma(0) f^\dagger_{A_1} = \gamma
\eqn\ArbLGequiv
$$
Although the diagonal-Coulomb gauge eliminates the continuous
gauge degrees of freedom, there still remain a set of discrete
transformations. These are implemented by gauge functions
satisfying the boundary conditions of eq. (\gaugeBC):
$$\eqalign{
&\Omega(t,x)_{jk}=\delta_{jk}~\exp\left\{ i\Bigl( \omega_j(t)+
{2\pi n_jx\over L}
+{2\pi m x\over L} \Bigr) \right\} \cr
&\qquad\sum_{j=1}^N \omega_j(t)=0 ~~[{\rm mod}~ 2\pi];~~~
  n_j \in {\bf Z},~~\sum_{j=1}^N n_j =0 ~~[{\rm mod}~ 2\pi],\cr
&~~~~~~~~~~m = {\ell\over N}, \qquad (\ell=1,\cdots,N-1), \cr}
\eqn\Weyla
$$
and the $N!-1 $ constant gauge transformations of the form
$$
\Omega =  \left(\matrix{
1&~&~&~&~\cr
{}~&0&-1&~&~\cr
{}~&1&0&~&~\cr
{}~&~&~&\ddots&~\cr
{}~&~&~&~&1}\right).
\eqn\Weylb
$$
which with the identity form the Weyl group of reflections $S$.
Under (\Weyla)
$$
\theta_j(t) \rightarrow \theta_j(t)+2\pi n_j+2\pi m,
\eqn\intotorus
$$
while under (\Weylb)
$$
\theta_i(t) \leftrightarrow \theta_j(t).
\eqn\Weylflip
$$

The $n_i$ term of the transformation (\intotorus) of represents the
$SU(N)$ periodicity of the maximal torus while the second term
reflects the $\Z_N$ symmetry. These residual (Gribov) gauge
transformations make the identifications needed to compactify the
configuration space $\Theta~(\sim \R^{N-1})$ into the maximal torus of
$SU(N)$ and thus are crucial in constructing the correct topology of
$Q_r$ [\JHAmstlat,\LSb].

The transformations (\Weylflip) produce the identifications which give
the configuration space it's orbifold structure, with the orbifold
singularities occuring on hyperplanes where $\theta_i = \theta_j$ for
any $i,j$. The freedom represented by the $\omega_j(t)$'s is used to
eliminate the $A_0$ components to which we now turn.

The Euler-Lagrange equations of motion for this model are
$$\eqalign{
\d_0 E +ig[A_0,E] &= 0  \cr
\d_1 E+ig[A_1,E] &= 0. \cr}
\eqn\EOM
$$
Since $A_0$ is periodic in $x$
$$\eqalign{
&A_0(t,x)_{ij} =\sum_n a_n(t)_{ij}~e^{2\pi inx/L}  \cr
&E(t,x)_{ij} = \delta_{ij} {1\over gL} \dot\theta_j
-{i\over L} \sum_n (2\pi n+\theta_i-\theta_j) a_n(t)_{ij}
e^{2\pi inx/L}.\cr}
\eqn\AzeroFourier
$$
Due to the second equation of (\EOM) and the gauge condition
(\diagCoulombdef)
$$D_1^2 E = D_1^2(\Theta) a_{n,ij}
= ( 2\pi n+\theta_i-\theta_j)^2~a_n(t)_{ij}=0.
\eqn\Azeromode
$$
Thus for general $\theta_i$'s, (\Azeromode) shows that
$a_n(t)_{ij}=0 ~(i\not= j)$, and the diagonal components can be gauged
away by choosing the $\omega_i(t)$'s in (\Weyla) appropriately.  In
general some of $a_n(t)_{ij} ~(i \not= j)$ may be nonvanishing when
$A_1$ lies on an orbifold singularity in the configuration space.
However eq. (\Azeromode) shows that these isolated configurations
do not contribute to $E(t,x)$. In other words, all
configurations are gauge equivalent to $A_0 =0$. Path integral
considerations leading to the same form were given in [\HH]. See
also [\GmodG].

In this form the theory is readily quantized since the $\theta_i$'s
are simply the coordinates of a point particle on the maximal torus
and the whole model reduces to a quantum mechanics problem,
albeit up to issues about quantization on orbifolds. In terms of the
coordinates on $T_{SU(N)}$, the Lagrangian is
$$
L~=\int_0^L dx~{\cal L}={1 \over g^2L}\sum_{k=1}^N \dot \theta_k^2
={1 \over g^2L}\Bigl( \sum_{k=1}^{N-1} \dot \theta_k^2
+ (\sum_{k=1}^{N-1} \dot \theta_k )^2 \Bigr).
\eqn\Loftheta
$$
Conjugate momenta to the $\theta_i$ are
$$
\pi_i={2 \over g^2L} \Bigl( \dot \theta_i + \sum_{j=1}^{N-1}
\dot \theta_j \Bigr) \equiv -i{\d\over \d\theta_i}
\eqn\Canonicalmom
$$
yielding the Hamiltonian
$$
H = -{g^2L\over 4} \Bigl\{ \sum_{j=1}^{N-1} {\d^2\over \d\theta_j^2}
-{1\over N}\Bigl( \sum_{j=1}^{N-1} {\d\over \d\theta_j} \Bigr) \Bigr\}.
\eqn\Hamth
$$

As shown above the configuration space of this model is an orbifold.
In such cases one usually quantizes on the simplest covering space of
the orbifold, demanding strict invariance of the wavefunction under
the discrete symmetries [\DowkerCone]. In our case the wavefunction
$\Psi(\theta_1,\dots,\theta_{N-1})$ will be a function on $T_{SU(N)}$,
periodic in each $\theta_i$ with period $2\pi$, and symmetric under
the Weyl reflections eq. (\Weylb). The wave function acquires a phase
$e^{i2\pi/N}$ to some power under a $\Z_N$ transformation as is
customary. This choice of periodicity makes the correspondence with
wavefunctions on the group manifold much more straightforward, as well
as simplifying the changes when fermions are added. Readers whose
needs require periodic wavefunctions under $\Z_N$ can easily restrict
the spectrum to the $N$th excitations, just as restricting to integer
$j$ states does for $SU(2) \rightarrow SO(3)$ representations.

In general we can construct wavefunctions from properly
symmetrized sums of phases,
$$
\Psi_{\{n\}}(\theta)  =\sum_{\{n\}}
c(n) ~e^{i(n_1\theta_1+ \cdots +n_{N-1}\theta_{N-1})}
\eqn\Psigen
$$
where $c(n)=c(n_1,\cdots,n_{N-1})$ and the $n$'s are some integers.
Since the wavefunction must be symmetric under the interchange of
$\theta_j$ and $\theta_k$ ($j,k=1,\cdots,N$), the $c(n)$'s must satisfy
$$\eqalign{
&c(n_1,n_2,\dots,n_{N-1})
  =c(n_1,\dots,n_j \leftrightarrow n_k,\dots,n_{N-1})~, \cr
&c(n)=c(m) ~,\qquad n_k=\cases{m_k-m_j,&for $k \not= j$ \cr
                             -m_j, &for $k=j$~. \cr} \cr}
\eqn\csyms
$$
Wavefunctions which satisfy (\csyms) are linear combinations of
$$
\Psi_n(\theta) = \sum_{\bar\theta} ~e^{in_1\bar\theta_1+ \cdots +
in_{N-1}\bar\theta_{N-1} }.
\eqn\Psisym
$$
Here $(\bar\theta_1,\cdots,\bar\theta_{N-1})$ are $N-1$
representatives made out of $\theta_k ~(k=1,\cdots,N)$ where
$\theta_N=-\sum_j^{N-1} \theta_j$.  The summation over $\bar\theta_j$
extends over all symmetric combinations in the indices
$k(=1,\cdots,N)$ of $\theta_k$.  The energy spectrum is given by a set
of integers $\{n_1,\cdots,n_{N-1}\}$~:
$$
E_{\{n\}} = {g^2L\over 4} \Bigl\{ \sum_j^{N-1}n_j^2 - {1\over N}~
\big(\sum_j^{N-1}n_j \big)^2 \Bigr\}.
\eqn\Egen
$$

The form of the wavefunction in (\Psigen) is rather specific to the
representation of $\Theta(t)$ in this Weyl basis, but we see that the
exponentials appearing in the sum (\Psigen) are simply Fourier modes a
complex function on the lattice defining $T^{N-1}$. To extract a
wavefunction for a general basis, note that this lattice is defined by
$$
\Gamma(n_1,n_2,\dots,n_{N-1}) = \prod_{i=1}^{N-1} z_i^{n_i}
\eqn\Tlatgen
$$
where $n_i \in \Z$, $z_i = e^{i{\bf h}_i}$ and the ${\bf h}_i$ are
the (suitably normalized) generators of the Cartan subalgebra in any
basis. $\Gamma$ is an element of the ``twist matrix'' which labels the
possible t'Hooft fluxes [\tHooftFlux,\Pierre,\Mark] of a
non-Abelian gauge field on a spatial torus with boundary conditions
$A_\mu(x+L_\nu) = U_\nu A_\mu(x) U_\nu^\dagger$ via the cocycle
condition $$ U_\mu(x_\nu+L_\nu) U_\nu(x_\mu) U_\mu(x_\nu)^\dagger
U_\nu(x_\mu+L_\mu)^\dagger = \Lambda_{\mu\nu}(n_1,n_2,\dots,n_{N-1}).
\eqn\thooftflux
$$
A Fourier mode on this lattice is then
$$
\Psi_{\{n\}}(y) = \prod_{i=1}^{N-1} e^{i n_i y_i}
\eqn\fouriery
$$
in terms of general coordinates $y_i$ of the maximal torus; $T_{SU(N)} =
\exp\{\sum_i y_i {\bf h}_i\}$.
Having these we construct a wavefunction by taking
the symmetric sum of these modes under Weyl transformations
$$
\Psi^{\rm sym}_n(y) = \sum_{\omega_{\rm eyl}} \Psi_n (S_\omega y_i)
\eqn\psisym
$$
in which $S_\omega$ implements the Weyl transformations.

As examples consider the groups $SU(2)$ and $SU(3)$. We have
straightforwardly from (\Hamth), for $SU(2)$:
$$\eqalign{
A_1&={1\over gL} \left( \matrix{\theta &~\cr
                          ~& -\theta \cr} \right) \qquad
H = -{g^2L\over 8} {\d^2\over \d\theta^2} \cr
E_n &= {g^2L\over 8}~n^2 \qquad
\Psi_n(\theta)= \sqrt{{2\over \pi}}\cos (n\theta) \cr
&~~~~~{\rm where} ~~~n=0,1,2,\dots\cr}
\eqn\sutwo
$$
which is simply the quantum mechanics of a particle on a circle,
whose wavefunctions are symmetric about $\theta = 0$.

The case of $SU(3)$ has a bit more internal structure.
In the Weyl basis, using (\Psigen) and (\csyms)
$$
A_1 ={1\over gL}\left(\matrix{\theta_{1}&~&~\cr
{}~&\theta_{2}&~\cr ~&~&-\theta_{1}-\theta_{2}\cr}\right) \qquad
H =-{g^{2}L\over 6} \Big({\d^2\over \d\theta_1^2}
-{\d\over \d\theta_1}{\d\over \d\theta_2} +{\d^2\over \d\theta_2^2}\Big)
\eqn\suthree
$$
$SU(3)$ states in this basis have energy
$$
E_{n_1n_2} = {g^2L\over 6}(n_1^2 -n_1 n_2 +n_2^2)
\equiv {g^2L\over 6}\varepsilon_{n_1n_2}.
\eqn\Weylspectrum
$$
Wavefunctions for some low-lying states are given in [\HH].

To illuminate the twist matrix discussion using a different basis, let
$$
{\bf h}_1 = \lambda_3 = \pmatrix{1&0&0\cr 0&-1&0\cr
0&0&0\cr},\qquad
{\bf h}_2 = \lambda_8
= {1\over \sqrt{3} } \pmatrix{1&0&0\cr 0&1&0\cr 0&0&-2\cr}.
\eqn\Cartanh
$$
The expansion $\Theta = \sum_i y_i h_i$ gives the relations
$$ \eqalign{
y_1 =&  {1\over 2}(\theta_1 - \theta_2)\cr
y_2 =&  {\sqrt{3}\over 2}(\theta_1 + \theta_2)\cr}
\eqn\yofth
$$
and the Hamiltonian (\Hamth) transforms to
$$
H = -{g^2L\over 8}\left( {\d^2\over \d y_1^2} + {\d^2\over \d y_2^2}\right)
\equiv -{g^2L\over 8}\Delta_{\bf h}
\eqn\LSHam
$$
which is the Hamiltonian obtained in [\LSa], up to normalization.
To use Fourier modes on the twist lattice $\Gamma$
as wavefunctions, recall that a general element of the twist matrix
(\thooftflux) for $SU(3)$ in the Gell-Mann basis (\Cartanh) is [\Mark]
$$\eqalign{
\Gamma(m_1,m_2) = (z_1^{m_1} z_2^{m_2}) &= \exp \left\{ 2\pi i
 \pmatrix{m_1+2m_2 &0&0\cr 0&m_1-m_2&0\cr  0&0&-2m_1-m_2\cr} \right\} \cr
\noalign{\kern 6pt}
&=\exp \left\{2\pi i\Big[{3\over 2}m_2 {\bf h_1}
 + {\sqrt{3}\over 2}(2m_1+ m_2) {\bf h_2} \Big] \right\} \cr}
\eqn\ZH
$$
where $m_1, m_2 \in \Z$.
A Fourier mode on $\Lambda$ in this basis is then
$$
\Psi_{m_1m_2}(y_1,y_2) = \exp \left\{ i m_2 y_1
 + i {1\over \sqrt{3}} (2m_1+ m_2) y_2 \right\}.
\eqn\psiy
$$
Applying the Hamiltonian (\LSHam) yields the energy
$$
E_{m_1,m_2} = {g^2L\over 6} (m_1^2 + m_1 m_2 + m_2^2)
\eqn\Cartanspectrum
$$
in exact agreement with eq. (\Weylspectrum). Since $m_1,m_2$ above and
$n_1,n_2$ in (\Egen) range over both positive and negative integers,
there is some state such that $E_{n_1,n_2} = E_{m_1,m_2}$ (or
simply use $\theta_2 \rightarrow -\theta_2$). The difference in the
energy for $n_1 = m_1, n_2 = m_2$ is due to the fact that the bases
defining the respective maximal tori have different fundamental Weyl
chambers.

\bigskip
\leftline{\bf B. Compact Formulation}
\bigskip

An alternative canonical quantization of this system was given by
Rajeev [\Rajeev] in the continuum, and is well known to lattice gauge
theorists [\KS,\Kogut,\Creutz,\Jan].  In this approach one does not
quantize the coordinates $A_1$, but instead maps the degrees of
freedom onto those of the Wilson loop $W$, ie. the system is viewed as
the quantum mechanics of a particle moving on the gauge group manifold
[\DowkerG]. In lattice gauge theory this is the definition of the
theory from the start, and the unreduced configuration space $Q_c$ is
just a large product of group manifolds, $G\times G\times \dots \times
G$.  After mapping the temporally gauge fixed coordinates to the
Wilson loop, Rajeev postulated the Laplace-Beltrami operator as the
Hamiltonian. In the lattice approach the same operator emerges without
gauge fixing and we review it's extraction and the natural projection
onto the gauge invariant sector of Hilbert space.

On the lattice, the Hamiltonian is extracted from the partition
function via the transfer matrix T,
$$\eqalign{
Z &= \int DU_\mu e^{S(U)} = {\rm Tr}~T^q \cr
  &= \prod_{i=0}^{q-1} \prod_{x,\mu}
\int dU_{x,\mu}(i)\la U_{x,\mu}(i+1)|\hat T |U_{x,\mu}(i)\ra \cr}
\eqn\latZT
$$
where $q$ is the number of time slices in the lattice.

The gauge freedom can be dealt with by first gauge fixing to the
reduced configuration space, or by projecting out gauge invariant
wavefunctions in Hilbert space after quantization of the unreduced
configuration space. To gauge fix a time slice of $n$ sites,
since a link transforms as $\Omega\cdot U_\mu (x) \rightarrow
G^\dagger(x)
U_\mu(x) G(x+\hat\mu)$, the gauge transformation, analogous to eq.
(\DiagCoulOmega)
$$
G_x =  \Big(\prod_{y=1}^{x-1} U_1(y)\Big)^\dagger \Lambda~\omega^{x-1}
\eqn\latCoulomb
$$
takes all spatial links into the Coulomb gauge $U_1(x) = \omega$,
where $\omega$ is the $n$th root of the Wilson loop,
$\omega^n = W = \prod_{x=1}^{n} U_1(x)$, and $\Lambda$ is arbitrary
(and can be used to diagonalize $W$).

Without gauge fixing, we proceed by noting that the transfer matrix
generally splits into potential and kinetic operators which
are functions of spatial coordinates and momenta
$$
\hat T = e^{-{1\over 2}V(x)} e^{K(\pi)} e^{-{1\over 2}V(x)}.
\eqn\latVKV
$$
$V(x)$ is a function of plaquettes with only space-like links,
of which there are none on a two dimensional lattice. Using
(\latZT) and the lattice action $S = -2/(g^2 a_x^2 a_t^2)\sum_x {\rm Re Tr}
[1 - U_\plq ]$ we have the matrix elements of $\hat T$
$$
\la U_1^\prime |\hat T |U_1\ra = \prod_x \int dU_0(x)
\exp\Big\{-{2\over g^2}{a_x\over a_t} {\rm Re Tr}
[1 - U_1(x) U_0(x+a_x) U^{\prime}_1(x)^\dagger U_0(x)^\dagger]\Big\}
\eqn\Tmatrixelm
$$
between time-like separated states in the coordinate basis $|U_1\ra$.
We further see that $\hat T$ factorizes into a kinetic
operator $T_K$ and a projection operator $P_\Omega$, which projects out
gauge invariant states from the Hilbert space $|U_1\ra$.

Let $\hat \Omega(U_0) |U_1\ra = | U_1^\Omega\ra$ where $U_1(x)^\Omega
= U_0^\dagger(x) U_1(x) U_0(x+a_x)$, then $\hat T  = \hat P_\Omega \hat
T_K$ where
$$
\hat P_\Omega = \prod_x \int dG(x) \hat \Omega(G)
\eqn\POmega
$$
and
$$
\la U_1^\prime |\hat T_K | U_1\ra = \prod_{x}
\exp\Big\{-{2\over g^2}{a_x\over a_t} {\rm Re Tr}
[1 - U_1(x) U^{\prime}_1(x)^\dagger]\Big\}
\eqn\TK
$$

It is instructive to observe the action of the projection operator
$\hat P_\Omega$ on wavefunctions $\la U | \Psi \ra = \Psi(U)$. In general
$$
\Psi(U) = \prod_x\sum_{j_x m_x n_x}\lambda_{j_x m_x n_x}
{\cal D}^{j_x}_{m_x n_x}[U_1(x)]
\eqn\PsiU
$$
using the completeness of the unitary irreducible representations
${\cal D}^j_{\alpha\beta}[G]$. Then (suppressing the $x$ dependence
of the indices)
$$\Psi(U^G) = \prod_x\sum_{jmn,ab}\lambda_{jmn}
{\cal D}^j_{ma}[G(x)^\dagger]{\cal D}^j_{ab}[U_1(x)]
{\cal D}^j_{bn}[G(x+a_x)]
\eqn\PsiUG
$$
{}From the orthogonality of the irreducible representations
we have
$$\eqalign{
\la U| \hat P_\Omega |\Psi\ra &= \prod_x\int dG(x) \Psi(U^G) \cr
&= \sum_{ji} \lambda^\prime_j {\cal D}^j_{ii}(\prod_x U_1(x)) \cr
&= \sum_j \lambda_j \chi_j(W) \cr}
\eqn\projchar
$$
so that $\hat P_\Omega$ projects out the wavefunction as a
series in conjugation invariant characters of the
Wilson loop as expected.

The Hamiltonian then follows from (\TK)
$$\eqalign{
\la W^\prime |\hat T_K | W\ra &=
   \exp\Big\{-{2\over g^2}{a_x\over a_t} {\rm Re Tr}
   [1 - W {W^\prime}^\dagger]\Big\} \cr
&= \int d\Gamma \exp\Big\{ i \gamma^i \cdot {\bf X}_i \Big\}
   \exp\Big\{-{2\over g^2}{a_x\over a_t} {\rm Re Tr}
   [1 - \Gamma]\Big\} \cr}
\eqn\TX
$$
where $\Gamma = \exp\{i \gamma \cdot \lambda\}$ and the ${\bf
X}_i$ are differential operators generating (left) translations on $G$.
As $a_t \rightarrow 0$, saddle point integration of (\TX) gives ($T \sim
e^{-a_t H}$)
$$
H = -\half {g^2\over a_x} {\bf X_i}^2
\eqn\HG
$$
showing the Hamiltonian to be proportional to
the quadratic Casimir operator of $G$.

Looking again at well known examples, the ${\bf X}_i$ for $SU(2)$ can
be written in Euler angles $W = W(\theta,\psi,\phi)$ say [\Vilenkin], as
$$\eqalign{
{\bf X}_1 &= \cos\psi {\d\over \d\theta}
            + {\sin\psi\over \sin\theta}{\d\over \d\phi}
            - \cot\theta~\sin\psi{\d\over \d\psi} \cr
{\bf X}_2 &= -\sin\psi{\d\over \d\theta}
             + {\cos\psi\over \sin\theta}{\d\over \d\phi}
             - \cot\theta~\cos\psi{\d\over \d\psi} \cr
{\bf X}_3 &= {\d\over \d\phi} \cr}
\eqn\Xs
$$
so that ${\bf X}^2$ is indeed the Laplace-Beltrami operator on
$SU(2)$. However since $P_\Omega$ projects out only radial wavefunctions
$\chi_j(\theta)$, we have
$$\eqalign{
H = -{g^2L\over 2}\Big\{&{\d^2\over \d\theta^2}
+ \cot\theta{\d\over \d\theta}\Big\} \equiv -\half \Delta_{LB}|_H
\qquad \chi_j(\theta) = {1\over \sqrt{2j+1}}{\sin[(j+\half)\theta]\over
\sin(\theta/2)} \cr
&E_j = {g^2L\over 2} j(j+1),
{}~~~~{\rm where}~~j = 0, \half, 1, {3\over 2},\dots \cr}
\eqn\SUtwoG
$$
The Casimir spectrum appearing above, $E_j \propto c_2^{SU(2)}$ is in
conflict though with that obtained in the previous section, $E_n
\propto n^2$ eq. (\sutwo), and a similar mismatch appears for $SU(3)$;
compare the non-compact spectrum eq. (\Cartanspectrum) $$
\varepsilon_{m_1m_2} = {1\over 6}(m_1^2 +m_1 m_2 +m_2^2)
\eqno{\rm \Cartanspectrum}
$$
with the quadratic Casimir of $SU(3)$
$$
c_2^{SU(3)}(\lambda,\mu) = {1\over 3}(\lambda^2 + \mu^2 + \lambda\mu)
        + \lambda + \mu
\eqn\QCsuthree
$$
The discrepancy persists for all $N$,
thus it appears that the two methods of quantization are in conflict
[\HH].

\secno=4  \meqno=1
\vskip 1cm
\leftline{\secrmb IV. Spectral Equivalence}
\bigskip

The previous section leaves us with the unpleasant result that
quantizing the theory in the group algebra $\g$ yields a different
spectrum than quantizing on the group manifold $G$. However the
essential feature, which emerged in both analyses, is that the
Hamiltonians are the radial parts of Laplacians acting on their
respective spaces. In fact, the projection of the radial parts of
differential operators on Lie groups and symmetric spaces has been
previously accounted by Berezin and Helgason [\Berezin,\Helgason],
among others. This projection has been used fruitfully by Dowker and
Schulman [\DowkerG,\Schulman] for instance, in computing the
propagator of a particle on a group manifold, allowing them to
diagonalize the radial Laplacian in the geodesic equation on $G$.

For Lie groups, the radial part $\Delta |_H$, of a Laplacian $\Delta$,
is a differential operator
over the Cartan subgroup $H \subset G$, invariant under automorphisms
of $H$:
$$
h\rightarrow g^\dagger h g \in H.
\eqn\autoh
$$
The adjoint action of $g \in G$ is such that
these motions form a finite set, the Weyl group $S$. Much of the
structure of $\Delta |_H$ is determined by its invariance under $S$.
The exponential map, $\exp: t \in \g \rightarrow g(t) = \exp(it)
\in G$
gives the canonical way of getting from the algebra
$\g$ to the group $G$. Then,
given a radial differential operator, $\Delta_{\bf h} =
P({\d\over \d t_1}, {\d\over \d t_2}, \dots, {\d\over \d t_{N-1}})$,
polynomial in ${\d\over \d t_i}$ on $\h$,
and invariant under (\autoh), we have
the corresponding radial Laplacian on $H$
$$\eqalign{
\Delta_H(P) &= {1\over J(t)}
P\Big({\d\over \d t_1}, {\d\over \d t_2}, \dots, {\d\over \d t_{N-1}}
\Big) J(t) \cr
& {\rm where}~~J(t) = \prod_{\alpha^+_i} \sin \big(
{<\alpha_i, t_i\alpha_i> \over 2} \big). \cr}
\eqn\mapping
$$
$\alpha^+_i$ are the positive roots of $\g$ and $J(t)$ enjoys the
relationship:~
$J^4 = \det g_{\mu\nu} \equiv |g|$ with the metric on $G$,
hence the Haar measure is $\prod_{\alpha^+} dt_\alpha J^2(t)$.

Evidently for
$$\Delta_{\bf h} \equiv
{\d^2\over \d t_1^2} + {\d^2\over \d t_2^2} +
\dots + {\d^2\over \d t^2_{N-1}}
= \sum_{i \in {\rm Cartan}} {\d^2\over \d t_i^2}
\eqn\deltah
$$
we obtain
$$\eqalign{
\Delta_H\Big(\Delta_{\bf h} \Big)
&= \Delta_{LB}|_r - R^2 \cr
\Delta_{LB} &= {1\over \sqrt{|g|}}\d_\mu (\sqrt{|g|} g^{\mu\nu} \d_\nu) \cr}
\eqn\htoH
$$
where $\Delta_{LB}|_r$ is the radial part of the Laplace-Beltrami
operator and $R$ is proportional to the scalar curvature of $G$.
For compact Lie groups $R$ is given by a familiar algebraic quantity
$$
R = \half \sum_{\alpha^+} \alpha_i~.
\eqn\Ralpha
$$
It should be remembered that the square of $R$ and
other scalar products involving the roots $\alpha_i$
are produced with the inner product on the root space via Killing form
$$
<a(h),b(h)> ~= \tr h_a h_b,
\eqn\Killingform
$$
where $a(h),b(h)$ are linear functionals of $h \in \h$ [\Cahn].

Here we see the source of the discrepancy in the spectra obtained
in the previous section. Since $\Delta_{\bf h}$ is a differential
operator on the flat manifold $\h$, to represent it faithfully
on $G$ we must subtract the curvature induced by the exponential
map when projecting $\Delta_{\bf h}$ to $G$. Thus the energy
eigenvalues are shifted up by the ``Casimir'' energy $R^2$.

In the non-compact $SU(2)$ example, eq. (\sutwo), neglecting now
the factors of $g^2L$
$$
H\Psi_n \sim -{1\over 8} \Delta_{\bf h} \Psi_n
= \half {n^2\over 4}\Psi_n = \half (j)^2 \Psi_n .
\eqn\sutwoj
$$
Using eq. (\htoH) to obtain the projected compact Hamiltonian,
$~\Delta_{\bf h}~-\kern-.45em\longrightarrow
{\kern-2.2em\raise1.3ex\hbox{exp}}~~\Delta_H:$
$$\eqalign{
H\chi_{_\lambda} \sim -\half\Delta_H \chi_{_\lambda}
= -\half (\Delta_{LB} - R^2)\chi_{_\lambda} &= \half
\big[ \lambda(\lambda+1) + {1\over 4} \big]\chi_{_\lambda} \cr
&= \half \big(\lambda+\half \big)^2\chi_{_\lambda} \cr
&= \half (j^\prime)^2 \chi_{_\lambda} \qquad \lambda = 0,\half,1,\dots\cr}
\eqn\sutwoshift
$$
using $\Big(\half\sum_{\alpha^+} \alpha_i\Big)^2 = {1\over 4}$,
for $SU(2)$. Thus including the curvature term gives exact agreement
of the spectra for $j > 0$. While comforting, this in itself is a rather
meager profit, being just an overall constant shift in the energy, however
we also observe a very interesting shift in the correspondence
between states.

We index states $\Psi_j$ on the algebra $\g$ now by $j = 0, \half,
1, \dots ={n\over 2}$, while on the group manifold $G$ the state of
corresponding energy to $j$, is $j^\prime = \lambda + R$ with
wavefunction $\chi_{_\lambda}$. Hence we have the correspondence
$$\eqalign{
{\rm on}~~\g = {\bf su(2)}:~~~~&~~~~~~~~~~~~~{\rm on}~~G = {\bf SU(2)}:\cr
j~~ &\longleftrightarrow ~~j - R \cr
\Psi_{j}~~ &\longleftrightarrow ~~ \chi_{j-R} \cr
-\half \Delta_{\bf h} \Psi_{j} = \half (j)^2\Psi_j~~ &\Longleftrightarrow
{}~~-\half \Delta_H \chi_{j-R} = -\half (\Delta_{LB} - R^2) \chi_{j-R} \cr
&~~~~~~~~~~~~~~~~~~~~~~~~~= \half (j)^2 \chi_{j-R} \cr
E_{\Psi_j} = {j^2\over 2} &~~=~~ {c_2(j-R) + R^2\over 2} = E_{\chi_{j-R}} \cr}
\eqn\wavefuncor
$$
An important consequence is that the ground states, $\Psi_0$ and
$\chi_0$ of these two quantizations are different, since $\chi_0
\leftrightarrow \Psi_\half$. The ground state of \g, $\Psi_0$,
corresponds to the unusual $\chi_{-\half}$ state to which we will
return in a moment. This structure is shown in Figure 1.
\bigskip
\vbox{ \epsfxsize= 10cm
\epsffile[150 400 460 540]{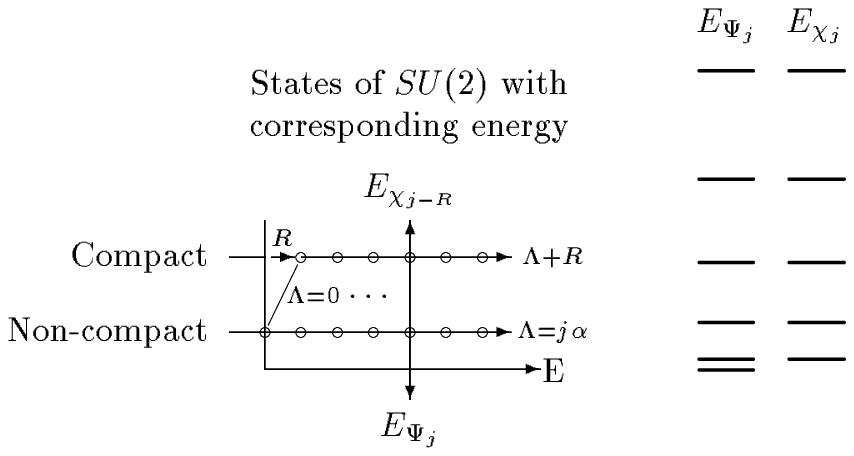}
\centerline{Correspondence of compact and non-compact spectra}
\centerline{and the shift of ground state for $SU(2)$.}
\centerline{------ Figure 1 ------} \bigskip}

Generalizing to $SU(3)$ and beyond, recall that [\Cahn]
$$
R_{SU(3)} = \half \sum_{\alpha^+} \alpha_i = \half  \alpha_1 +
\half  \alpha_2 + \half (\alpha_1 + \alpha_2)
=  \alpha_1 +  \alpha_2
\eqn\suthreeR
$$
An $SU(3)$ representation is labeled by a (highest) weight in
the fundamental Weyl chamber of the weight lattice and can be written
$$
\Lambda(\lambda,\mu) = {1\over 3}(2\lambda + \mu)\alpha_1 +
                       {1\over 3}(\lambda + 2\mu)\alpha_2
\eqn\suthreeLambda
$$
We normalize the root lattice such that
$$\eqalign{
<\alpha_1,\alpha_1> &= <\alpha_2,\alpha_2> = 1 \cr
<\alpha_1,\alpha_2> &= -\half ~. \cr }
\eqn\rootinners
$$

\vbox{
\epsfxsize= 10cm
\epsffile[150 360 450 520]{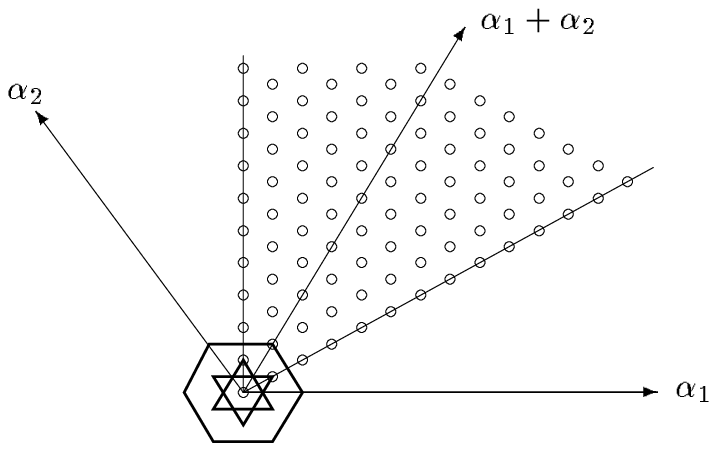}
\centerline{The highest weights of representations
$\Lambda(\lambda,\mu)$ in the fundamental Weyl}
\centerline{chamber of $SU(3)$. The {\bf 3},
{\bf 3$^*$}, and {\bf 8} are displayed.}
\centerline{------ Figure 2 ------}
\bigskip}
For $SU(2)$ we saw,
$$
E^\Psi_{j+R} = \half \big(j + \half\big)^2
= \half \big(j\alpha+R\big)^2 = E^\chi_j
= \half \big(c_2(j) + R^2\big)
\eqn\Eshiftsutwo
$$
Then using eq. (\Cartanspectrum) and
$\Lambda(\lambda,\mu) + R = \Lambda(\lambda+1,\mu+1)$ we find
the same behavior for $SU(3)$
$$\eqalign{
E^\Psi_{\Lambda+R}
&= {1\over 6} \big[ (\lambda+1)^2 + (\lambda+1)(\mu+1) +
(\mu+1)^2 \big] \cr
&= {1\over 6}(\lambda^2 + \lambda\mu + \mu^2) + \half(\lambda + \mu) +
\half \cr
& = \half \Big(c^{SU(3)}_2(\lambda,\mu) + R^2\Big)~. \cr}
\eqn\suthreeshift
$$
The displacement of the Weyl chamber boundary states which are
missing in compact quantization is shown in Figure 3.

\vbox{
\epsfxsize= 10cm
\epsffile[150 340 510 530]{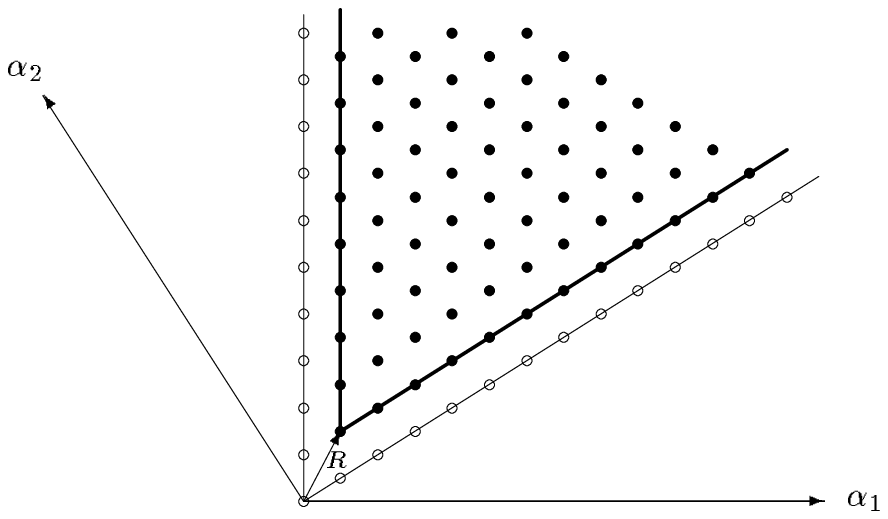}
\centerline{The shift of lowest energy states from non-compact to compact
quantization for $SU(3)$}
\centerline{------ Figure 3 ------}
\bigskip}
Clearly this structure maintains for all $G$, since the
quadratic Casimir of the state labeled by $\Lambda$ is given by
$$
c^G_2(\Lambda) = <\Lambda + R, \Lambda + R> - <R,R>
\eqn\generalCasimir
$$
so that $\half (c^G_2 + R^2)$ always yields the energy
$\half <\Lambda + R, \Lambda + R>$ for the Fourier mode $\Psi_{\Lambda
+R}$ on \g.  Compare for example, $\half <\Lambda, \Lambda>$ with
the non-compact spectrum of $SU(3)$ in eq. (\Cartanspectrum):
$$\eqalign{
\half <\Lambda,\Lambda> &= \Big({1\over 3}(2\lambda + \mu)\alpha_1 +
                      {1\over 3}(\lambda + 2\mu)\alpha_2\Big)^2 \cr
                  &= {1\over 6}(\lambda^2 + \lambda\mu + \mu^2)}
\eqn\Lambdasq
$$

\secno=5  \meqno=1
\vskip 2cm
\leftline{\secrmb V. State Mapping}
\bigskip

In the previous section we saw that the spectral decompositions
of $\Delta_{\bf h}$ and $\Delta_H$ are identified, revealing a shift
in the corresponding wavefunctions, induced by the curvature of $G$.
We now examine these wavefunctions more closely.

The $SU(2)$ Hamiltonian on the algebra \g, was simply proportional to
${\d^2\over \d\theta^2}$ with solutions $\Psi_n = \cos(n\theta)$,
$\sin(n\theta)$: the odd and even solutions under the Weyl reflection
$S: \theta \rightarrow -\theta$.
Invariance selects $\Psi^+_n = \cos(n\theta)$.

On $G$, the wavefunction $\chi_{_\lambda}$ is also Weyl invariant
$$
\chi^+_\lambda(\theta)
= {\sin[(\lambda + \half)\theta]\over J(\theta)}
={\sin[(\lambda + \half)\theta]\over \sin({\theta\over 2})}
\eqn\chioverJ
$$
since both numerator and denominator are odd. The numerator
corresponds, through eq. (\mapping), to the odd wavefunction
on $\g$: $J(\theta)\chi_\lambda = \Psi^-_{\lambda +R}
 = \sin[(\lambda + \half)\theta]$~ (recall that $\theta \rightarrow
{\theta\over 2}$ from $\g \rightarrow G$ in our notation).
In passing, notice that the Weyl-odd function, using the even
eigenfunction on $\g$
$$
\chi^-_\lambda = {\cos[(\lambda + \half)\theta]\over J(\theta)}
\eqn\oddonG
$$
is also an eigenfunction of $\Delta_{LB}$ with the same Casimir
eigenvalue $-\lambda(\lambda + 1)$. However it cannot be globally
defined as a function on $SU(2)$, since it is impossible to have
an odd function of polar coordinates, which is equatorially constant.

This parity shifting between wave functions on $\g$
and $G$ is a general feature of all semi-simple Lie groups [\Berezin].
The denominator appearing in the Weyl character formula is always
odd, due to its geometrical nature. In general it is defined as
$$\eqalign{
J(t_1,t_2,\dots,t_n)
&= \prod_{\alpha^+}\Big\{e^{i\half <\alpha, t>}
- e^{-i\half <\alpha, t>}\Big\}\cr
&= \sum_\omega (\det S_\omega) e^{i<R,S_\omega t>}
\qquad t = \sum_i t_i\alpha_i \cr}
\eqn\Weyldenom
$$
Functions such as $J(t)$, transforming as~ $S_\omega J(t) = (\det
S_\omega) J(t)$~ are called alternating. The Weyl character formula
$$
\chi_{_\Lambda}(t_1,\dots,t_n) = {\sum_\omega (\det S_\omega)
e^{i< \Lambda +R,S_\omega t>} \over
\sum_\omega (\det S_\omega) e^{i<R,S_\omega t>}}
\eqn\Weylchar
$$
uses an alternating numerator, the odd Fourier mode $\Lambda +
R$, divided by $J(t)$ to produce an invariant character.  In contrast,
the general form of an invariant wavefunction on $\g$ is
$$
\Psi_{\Lambda}(t_1,\dots,t_n)
= \sum_\omega e^{i< \Lambda,S_\omega t>}
\eqn\PsiLambdaprime
$$
using the even mode $\Lambda$.

What then of the state corresponding to $\Psi^+_0$ on $\g$ when
projected to $G$?
Returning to $SU(2)$ we find in fact, that the Weyl-odd state,
$\chi^-_{-\half}$, corresponding to $\Psi_0$
is an eigenfunction of $\Delta_{LB}$ with eigenvalue $R^2$
$$\eqalign{
\chi_{-\half} &= {1\over J(\theta)} = {1\over \sin({\theta\over 2})} \cr
\Delta_{LB} \chi_{-\half} &= {1\over 4} \chi_{-\half} = R^2\chi_{-\half} \cr
\Longrightarrow~\Delta_H\chi_{-\half}
&= -\big(\Delta_{LB} - R^2\big)\chi_{-\half} = 0 \cr}
\eqn\chihalf
$$
so that the projected zero mode, $\Psi_0 \longrightarrow
\chi^-_{-\half}$, is a zero mode of the projected Hamiltonian
$\Delta_{\bf h} \longrightarrow \Delta_H$, but is not however,
a globally defined function on $G$.

It is possible to construct a Weyl-even zero mode, since both $1$ and
$\theta$ belong to the kernel of $\Delta_{\bf h}$.
$$
\chi^+_{-\half} = {\theta\over \sin({\theta\over 2})}
\eqn\chiRuse
$$
also satisfies $\Delta_H \chi^+_{-\half} = 0$.  This state is apparently
related to Ruse' invariant $\rho$ of a harmonic manifold
[\Ruse,\Dowkerpath], which is defined by
$$
\rho(x_0,x) = {\sqrt{|g_0||g|}\over J(x_0,x)},\qquad
J(x_0,x) = {\d^2\Omega\over \d x_0\d x},
\qquad \Omega = \half g_{\mu\nu}x^\mu_0 x^\nu = \half s^2
\eqn\Rusedef
$$
For spaces of constant curvature
$$\eqalign{
R_{ijkl} &= \kappa (g_{ik}g_{jl} - g_{il}g_{jk}) \cr
\rho(s) &= {\sin^2(\sqrt{\kappa} s)\over \kappa s^2} \cr
\Delta_{LB}~ \rho^{-\half} &= \kappa~ \rho^{-\half} \cr}
\eqn\RuseconstR
$$

This state does not belong to the Hilbert space of $\Delta_H$,
although it is square integrable on $SU(2)$. While having the correct
Weyl symmetries, it is not periodic in $\theta$.  Dowker has dubbed
this state the "zero representation", in which all group elements are
represented by ${\bf 0}$ [\Dowkerpath]. The dimension of a representation
$\Lambda$ is
$$
{\rm dim~}\Lambda = \sum_{\alpha^+}{<\alpha_i, \Lambda + R>\over
<\alpha_i,R>}.
\eqn\Weyldim
$$
Hence, using the correspondence found above between the states
$\Lambda_g \leftrightarrow \Lambda_G - R$, we see that any state
$\Psi_\Lambda$ which lies on the boundary of a Weyl chamber will
project to such a zero-dimensional character in this sense, and thus
not appear in the compact spectrum. In non-compact quantization these
states have a constant (zero mode) wavefunction along some sub-torus of
the maximal torus.

\secno=6  \meqno=1
\vskip 1cm
\leftline{\secrmb VI. Conclusions}
\bigskip

We have seen that straightforward canonical quantization of two
dimensional Yang-Mills theory leads to two definitions of the quantum
theory, depending on the topology we allow for the configuration
space.  On the one hand we can quantize the field $A_\mu$, as a
quantum theory on the Cartan subalgebra $\h \subset \g$ subject to
the periodicities of the maximal torus. Thus the configuration space
is $\h /\Z^{N-1} = T^{N-1}$ (up to the identification of Weyl
reflections). On the other hand, we can map the gauge field to the
group manifold $G$ itself and quantize the system there, restricting
the wavefunctions to class functions of $G$, ie. they are again only
functions of the coordinates of the maximal torus $T^{N-1} \subset G$.

Two slight subtleties in relating the quantizations are that
$G$ has constant curvature while $\g$ is flat so that, relative
to $\g$, the spectrum on $G$ is shifted by a constant ``Casimir
energy'' (in a conspiracy of terms) as usually happens upon
compactification. More importantly the wavefunctions in compact
quantization must be globally defined functions on $G$, whereas
non-compact wavefunctions live only on the maximal torus.

Keeping track of the correspondence of states, we found that the zero
modes of the non-compact Hamiltonian disappear when the Hamiltonian is
mapped to $G$, due to the new topology confronting wavefunctions. By
zero mode here is meant any wavefunction $\Psi_{n_1 n_2 \dots 0 \dots
n_{N-1}}(\theta_i)$ which is constant around one of the circles of the
maximal torus, so that it is a zero mode of the Laplacian of that
circle.  In the root plane these states lie on the boundaries of the
Weyl chamber. This effect is similar to the reason the zero mode of
the Dirac operator is absent under compactification of $\R^2$ to $S^2$
for instance [\Stone], where $(i\gamma^\mu\d_\mu)^2 = -\nabla^2 +
{1\over 4}R$, inducing the same type of shift as in (\htoH). Thus the
primary observation is that the two quantizations have different
ground states. Two areas of relevance for these observations certainly
come to mind, namely: lattice gauge theory and the recently discovered
sum over maps representation of the 2D-QCD partition function
[\GrossTaylor,\Minahan].

Clearly the compact lattice formulation of 2DYM will reproduce the
compact quantization, more or less by definition. However the fact
that the continuum theory has states of lower energy which cannot be
seen even in the exact solution and classical continuum limit of the
lattice model is rather unexpected. In the canonical formalism we see
that the topology of the configuration space is experienced by the
wavefunctions, which of course are global objects. As usual, this
topology is most severely felt by the zero modes of the Laplacian.

Intricately woven into the above analysis is gauge fixing, which
allows us to identify exactly the coordinates of the physical
configuration space and quantize {\it only} these. On the lattice,
without gauge fixing, wavefunctions live on the unreduced
configuration space $Q = G^{n_{\rm links}}$, hence all degrees of
freedom are quantized.  Perhaps by a strong form of gauge fixing we
can restrict the quantization on the lattice to the physical
configuration space so as to reproduce the non-compact results.

Strictly speaking we have analyzed here essentially the one plaquette
model for the dynamics of a single link [\Kogut,\JS].  If these two
inequivalent compact and non-compact quantizations persist to higher
dimensional gauge theories, the implications could be very far
reaching, however this is rather difficult to analyze since the
Hamiltonian is then not just a differential operator in the radial
coordinates of the maximal torus. For a non-radial Hamiltonian mapping
the Lie algebra to $G$ is often necessitated in order to make residual
gauge identifications of configuration space possible. This is the
case even in two dimensions for instance. Using only the Coulomb gauge
condition $\d_1 A_1 = 0$ without diagonalizing $A_1$, produces a
complicated non-radial Hamiltonian with residual gauge symmetries.
Similar shifts in spectra have been encountered upon interchanging the
order of gauge fixing and quantization in simple models [\Renata] thus
the study of the gauge fixing --- quantization process, while at the
same time compactifying the configuration space as in lattice gauge
theory demands further study. A recent BRST quantization of 2DYM has
revealed more surprises as an anomaly develops in the Kac-Moody
algebra of the constraints for certain polarizations of phase
space [\KalauBRST].

As mentioned above, it was recently shown that the partition function
of 2DYM on a Riemann surface, $M_G$ of genus $G$, computed in the
continuum limit of the compact lattice formulation and
expanded in ${1\over N}$ can be
represented as $N \rightarrow \infty$, as the sum over homotopicly
distinct maps from a Riemannian worldsheet, $W_g$ of genus $g$, to the
target space $M_G$.
$$\eqalign{
Z(M,e) &= \int {\cal D}A_\mu
\exp\Big\{ -\half\int_M d^2x\sqrt{g}\tr F_{\mu\nu}F^{\mu\nu} \Big\}\cr
&= \sum_{{\rm reps}~ \Lambda} d_\Lambda^{2-2G}\exp\Big\{
  -\half e^2 A c_2(\Lambda)\Big\} \cr
&\longrightarrow {\kern-2.5em\lower2.0ex\hbox{$^{N\rightarrow\infty}$}}
  \sum_{g=0}^\infty \sum^\infty_{n=1}\sum_i N^{2-2g}
  \omega_{g,G}^{n,i}(e^2NA)^i\exp\big[ -n{e^2A\over 2N} \big] \cr}
\eqn\sumofmaps
$$
where $e$ is the coupling constant of the Yang-Mills fields, and
$\omega_{g,G}^{n,i}$ counts the number of topologically distinct,
smooth maps from $W_g$ to $M_G$, with winding $n$, and $i$ branch points.

Two very interesting features of this string theory representation are
the following. First, no degenerate maps occur in which the worldsheet
fails to cover the target space at least once, ie.  there are no maps
of the entire worldsheet to a point or a Wilson loop on $M_G$.
Furthermore only one smooth map per topological class is summed, as
opposed to the usual string path integral in which {\it all} possible
maps are integrated over, including folded maps in the same homotopy
class.

The canonical treatment above reveals that the quantum states of 2DYM
are essentially the Fourier modes on the maximal torus of $G$, with
non-compact quantization picking the even cosine modes, while compact
quantization chooses the odd sine modes and divides them by the Weyl
denominator $J(t)$. As such, these Fourier modes provide one
representative map from some sub-torus of the maximal torus to the
cylindrical spacetime, for each winding number $n$. This hints that the
worldsheets are really sub-tori of the maximal torus.
In terms of this observation the missing states
in the compact quantization which are constant Fourier modes along
some sub-torus, would be the zero winding maps absent in the string
representation of 2DYM.

Expressed differently, the above analysis shows that (at least for
genus $G=1$) the partition function (\sumofmaps) should be modified to
$$
Z = \sum^\infty_{{\rm reps}~ \Lambda = -R} d_\Lambda^{2-2G}
  e^{ -\half e^2 A[c_2(\Lambda) + R^2]}
\eqn\Zmod
$$
picking up an overall factor $e^{-\half e^2 A R^2}$ due to compact
quantization. While this is a trivial factor, the real
issue is the range of the sum over representations. To express
the quantization of $A_\mu$ via compact quantization, we must include
the ``zero representations'', $\chi_{-R}$, which are present in
non-compact quantization and include the ground state of the
theory defined by ${\cal L} = \half \tr \int dx^2 F_{\mu\nu}^2$

\vskip 1cm
\leftline{\secrmb Acknowledgements}
\bigskip

I wish to thank Y. Hosotani, J. Smit, R. Dijkgraaf, Ph. de Forcrand,
D. Birmingham, and P. van Baal for stimulating and thoughtful
discussions. This research is supported by the Stichting voor Fundamenteel
Onderzoek der Materie (FOM).

  \vfill\eject\immediate\closeout\reffile
  \centerline{{\bf References}}\bigskip\frenchspacing%
  \input refs.tmp\vfill\eject\nonfrenchspacing

\newwrite\advert
\immediate\write\advert{Plain TeX---no muss, no fuss...}

\bye